\documentclass[conference]{IEEEtran}
\IEEEoverridecommandlockouts
\usepackage[english = american]{csquotes} 
\usepackage{url,multirow,comment,hyperref,amsmath,graphicx,amsmath,subfigure,listings,tabularx,psfrag,wrapfig,cite,verbatim,array, algorithm, algorithmicx, algpseudocode}
\usepackage[
locale = DE 
]{siunitx} 
\usepackage{caption} 
\usepackage{xcolor,mathtools}
 
\def\BibTeX{{\rm B\kern-.05em{\sc i\kern-.025em b}\kern-.08em
    T\kern-.1667em\lower.7ex\hbox{E}\kern-.125emX}}
\begin{document}

\title{Power and Skew Reduction Using Resonant Energy Recycling in 14-nm FinFET Clocks \\
\thanks{This work was supported in part by the
Rezonent Inc. under Grant CORP-0061 and the UMBC Startup grant.}
}

\author{\IEEEauthorblockN{
Dhandeep Challagundla, Mehedi Galib}
\IEEEauthorblockA{\textit{Computer Science and Electrical Engineering} \\
\textit{University of Maryland Baltimore County}\\
Baltimore, Maryland, USA \\
\{vd58139, mgalib1\}@umbc.edu}
\and
\IEEEauthorblockN{
Ignatius Bezzam}
\IEEEauthorblockA{\textit{IC Design Group} \\
\textit{Rezonent Inc.}\\
Milpitas, USA \\
i@rezonent.us}
\and
\IEEEauthorblockN{
Riadul Islam}
\IEEEauthorblockA{\textit{Computer Science and Electrical Engineering} \\
\textit{University of Maryland Baltimore County}\\
Baltimore, Maryland, USA \\
riaduli@umbc.edu}
}

\maketitle

\begin{abstract}
As the demand for high-performance microprocessors increases, the circuit complexity and the rate of data transfer increases resulting in higher power consumption. We propose a clocking architecture that uses a series LC resonance and inductor matching technique to address this bottleneck. By employing pulsed resonance, the switching power dissipated is recycled back. The inductor matching technique aids in reducing the skew, increasing the robustness of the clock network. This new resonant architecture saves over 43\% power and 91\% skew clocking a range of 1--5~$GHz$, compared to a conventional primary-secondary flip-flop-based CMOS architecture. 
\end{abstract}

\section{Introduction}
Power consumption is one of the major problems faced in the high-performance microprocessor industry~\cite{Fischer:2011, Cunningham:2021,kumardesign}. The need for an increase in performance has steered the operating frequencies higher, resulting in an increased complexity among the microprocessor designs~\cite{Cunningham:2021, kumar2021novel, khan2019migration}. This higher power led designers to constantly come up with innovative techniques to reduce the power while trying to meet all the design constraints that impact the performance~\cite{jeong2018sense,touil2020design,cet2018review,8370498}.
A significant portion of dynamic power consumed in a high-frequency design is due to the switching activity in the clock network~\cite{rabaey_2009}.
To address this, several low power techniques such as dynamic voltage and frequency scaling (DVFS)~\cite{Nowka:2002}, clock gating~\cite{Tirumalashetty:2007} and LC resonant clocking{~\cite{Rahman:2018,Bezzam:2015,Fuketa:2014,Islam:2018, Lin:2015}, current-mode clocking~\cite{Islam:2017, Guthaus:2014, Islam:2015} are commonly used. Among them, inductor-based LC resonant clocking techniques have great potential to save switching power due to their constant phase and magnitude. 
\begin{figure}[t!]
	\centerline{\includegraphics[width = 0.33\textwidth]{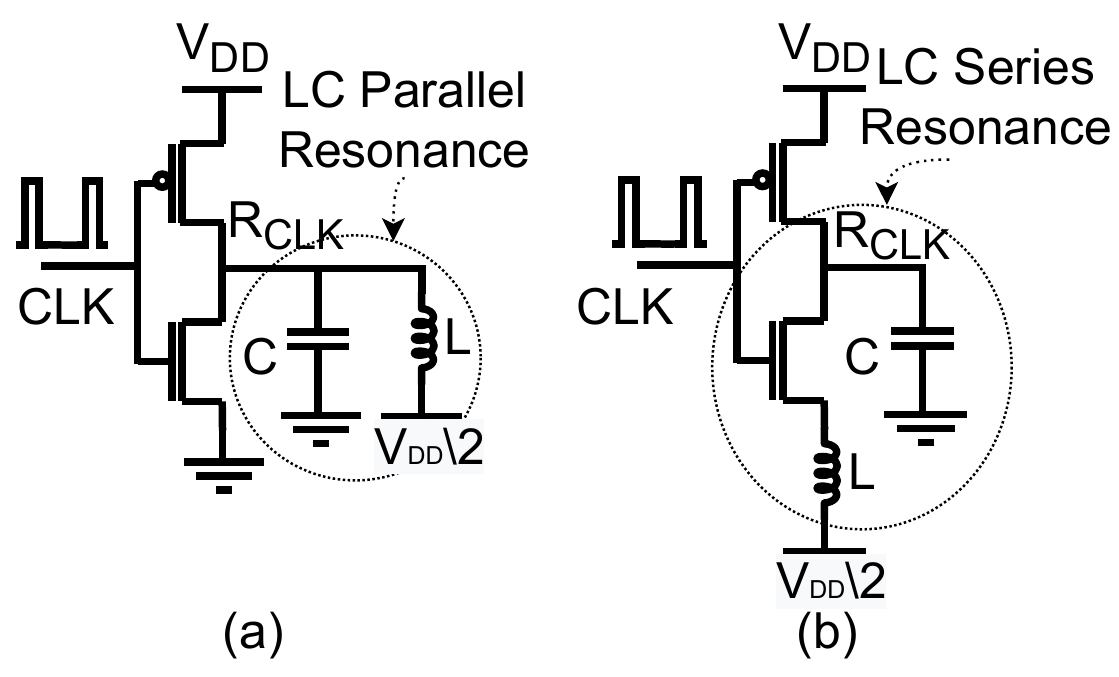}}
	\caption {LC resonance topologies to reduce dynamic power consumption (a) series resonance topology can address wide frequency band (b) parallel resonance topology can address very narrow frequency band.} 
	\label{fig:par_vs_ser_rez}
	\vspace{-0.5cm}
\end{figure}
There are several LC resonant techniques, such as parallel resonance~\cite{Rahman:2018} (please see Fig.~\ref{fig:par_vs_ser_rez}(a)), intermittent resonance~\cite{Fuketa:2014} and series resonance~\cite{Bezzam:2015, Bezzam:2021}. However, most of the LC resonant techniques suffer from a limitation of narrow frequency band and high skew. Moreover, most of the industry-standard electronic design automation (EDA) tools do not explicitly support integrating LC resonance in the clock tree architecture. Additionally, designing a resonant clock architecture requires the designer to have multiple domain expertise due to the non-linear behavior of inductors.

To overcome the narrow frequency band, researchers utilize series resonance techniques~\cite{Bezzam:2015}, as shown in Fig.~\ref{fig:par_vs_ser_rez}(b). This approach uses an inductor placed in the discharge path to store the dissipated energy in the form of a magnetic field. This energy is recycled in the next rising clock edge.

To enable a resonant clock architecture, we need resonant flip-flops (FFs) for synchronous circuits. However, they occupy a substantial chip area consuming high power. 
Researchers proposed many low-power flip-flops, however, not suitable for resonant operations~\cite{Cai:2019,Stas:2018,Tsai:2018}. In this research, we propose several conventional register-based pulsed FFs suitable for series resonance and reduce the overall power consumption in the clock network.

Besides power, skew plays a critical role in enabling high-frequency operation. To reduce the skew generated by clock trees, we introduce the first inductor tuning technique to match the series resonance inductor with the load capacitance of the clock tree, which generates pulses with equal resonant frequencies. As the resonant frequency depends on the series inductor and the load capacitance, we generate a constant pulse-width for a wideband (WB) of input clock frequencies. Therefore, calibrating the inductors once for a clock tree architecture enable WB frequency operation. 

This paper proposes a clocking architecture to recycle the power dissipated by the synchronous elements using series resonance technology and improve clock skew by adapting the inductor tuning technique. In particular, the main contributions of this work are:
\begin{itemize}
\item An architecture to recycle the power dissipated by synchronous elements using series resonance.
\item A novel pulse generator with dual-rail booster using series resonance.
\item A set of pulsed register-based FFs that exploits the behavior of pulsed series resonance.
\item An inductor tuning technique to compensate for the skew in the clock tree architecture.
\end{itemize}

\section{Background}

In a traditional clocking method, half of the switching power is utilized to charge a capacitive node when the clock transitions from 0-to-1. The other half of switching power is dissipated in the discharge cycle when the clock transitions from 1-to-0. The pulsed series resonance (PSR) technique recycles this dissipated energy by placing an inductor in the discharging path.
LC resonant is most widely used among several energy recycling techniques as it precisely replicates conventional CMOS clocking. However, it suffers from a higher slew rate while demonstrating great savings in dynamic power consumption~\cite{Bezzam:2015}~\cite{Islam:2021}. 

Due to resonance, the free energy swing obtained as a result of recycling energy, is the difference between resonant high output~($V_{OH}$) and resonant low outupt~($V_{OL}$)~\cite{Bezzam:2015} can be expressed as
\begin{equation}
\label{energy_saved}
V_{OH}-V_{OL} = \frac{V_{DD}}{2}(1+e^{-\pi/Q}) - \frac{V_{DD}}{2}(1-e^{-\pi/2Q})
\end{equation}

where $Q$ is the quality factor of the inductor, which is given by~$Q=\sqrt{{L}/{(CR^2)}}$. We utilize an external power source to pull the output from $V_{OH}$ to $V_{DD}$. The resonant frequency at which the inductor resonates with the load capacitance can be expressed as $f_{RES}=\dfrac{1}{2\pi\sqrt{LC}}$.

In this work, we proposed several pulsed-type resonant FFs with resonant clock trees for PSR operation. Several prior works have focused on low-power FF designs. In~\cite{Stas:2018} an 18T FF was designed to achieve 40\% improvement in energy/cycle compared to primary-secondary FF (PSFF). However, to mitigate voltage degradation caused due to the non-complementary topology, the authors used a poly bias technique~\cite{Cai:2019}, which requires extra design effort. In~\cite{Cai:2019} an 18T single-phase clocked FF was designed for low power operation. It showed 68\% lower power consumption at 0.6V supply but had functionality issues when the voltage was scaled, as reported in~\cite{Shin:2020, You:2021}. For reliability and robust operation, we implement widely used traditional pulsed register-based and true single-phase clock (TSPC)-based FFs~\cite{Rabaey:2010}.

\section{Proposed Clock Architecture}
The proposed architecture comprises a pulse generator, clock drivers utilizing on-chip inductors, and pulsed registers, as shown in Fig.~\ref{fig:proposed_tree}. 
\begin{figure}[t!]
	\centerline{\includegraphics[width = 0.45\textwidth]{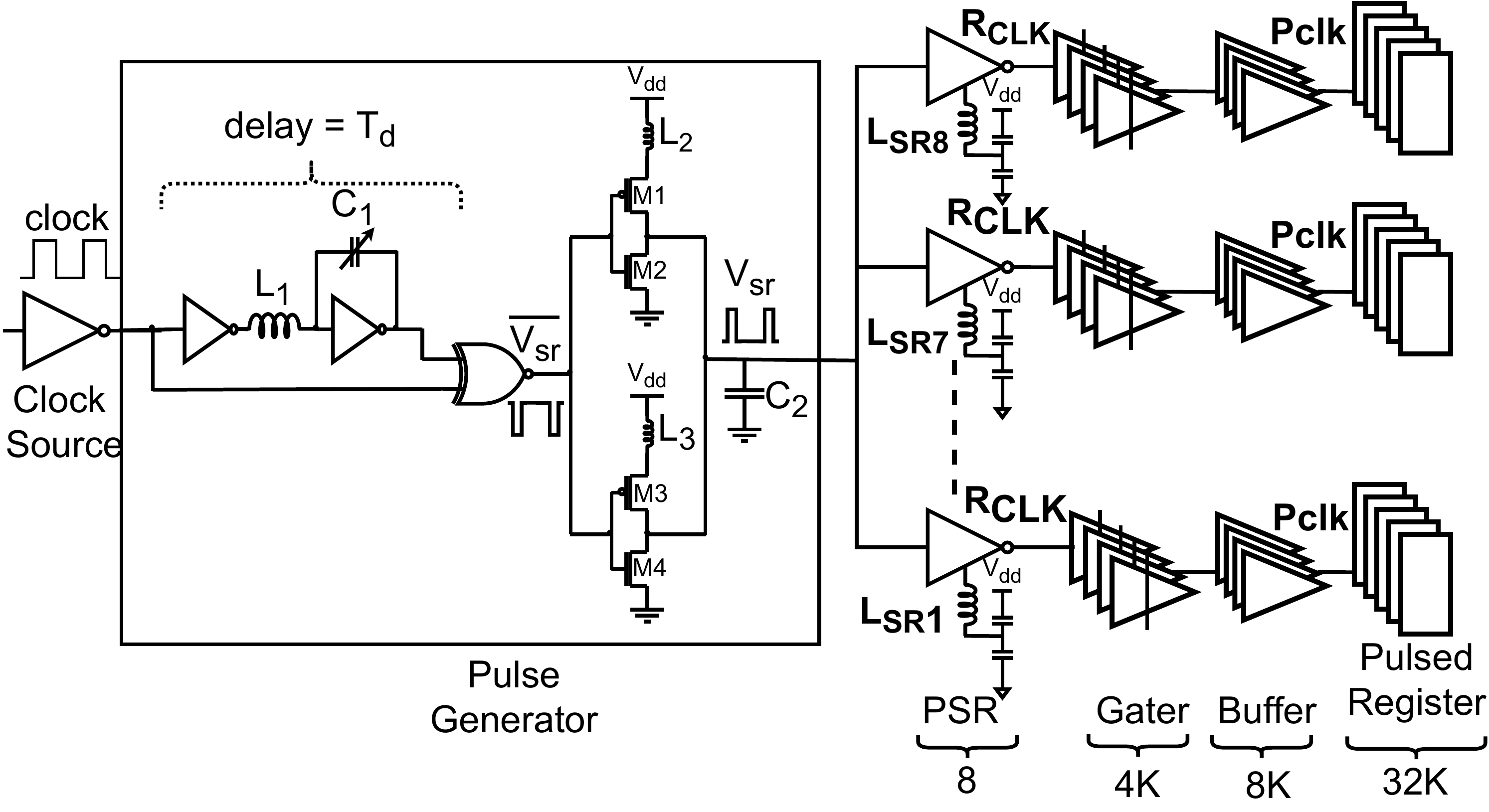}}
	\caption {The proposed wideband resonant clock tree architecture consists of a system clock source as the root, followed by a pulse generator, multiple PSR drivers with on-chip inductors, clock gaters, clock buffers, and finally, various sets of resonant pulsed FFs in the leaf nodes.} 
	\label{fig:proposed_tree}
	\vspace{-0.25cm}
\end{figure}

\subsection{Pulse Generator}
The pulse generator is depicted in Fig.~\ref{fig:proposed_tree} takes the input from the clock source and generates a pulse with boosted amplitude. 
The series inductor $L_1$ and the matching capacitance $C_1$ generate a delay of $T_d=\pi\sqrt{L_1C_1}$.  The clock and the delayed signal are fed into an XNOR gate to generate a pulse at both clock edges with a pulse width $T_d$. Now, a voltage doubler circuit is employed to invert the generated dual triggered pulse resulting in a boosted signal $V_{SR}$. The voltage doubler circuit uses the pulsed series resonance technique to generate a boosted signal. When the $\overline{V_{sr}}$ is low, the PMOS transistors $M_1$ and $M_3$ are ``ON,'' and the inductor resonates with the load capacitance $C_2$ and the additional PSR capacitance. For large load capacitances, the value of the series inductors is quite small. The inductor in the voltage doubler circuit can be adjusted according to the load of the pulse generator to produce a boosted signal $V_{SR}$. 
We use a dual-rail booster circuit to reduce the power consumed by the voltage doubler by decreasing the resistance of the pull-up network.

\subsection{Pulsed Series Resonance Driver}
The boosted signal $V_{SR}$ from the pulse generator stage is provided as the input to multiple PSR drivers to generate a pulse signal $R_{CLK}$. The inductors on the PSR drivers resonate with the capacitance of the tree to generate a pulse signal that is traversed through many levels of transmission-gate clock gaters and clock buffers. Since we provide a boosted $V_{SR}$ signal as the input, we obtain a rail-to-rail swing at the output of the PSR driver, and it improves the robustness of the design. Then, the output signal of the PSR driver $R_{CLK}$ is inverted and supplied to pulsed registers as the clock input signal $Pclk$. 
\begin{figure}[t!]
	\centerline{\includegraphics[width = 0.46\textwidth]{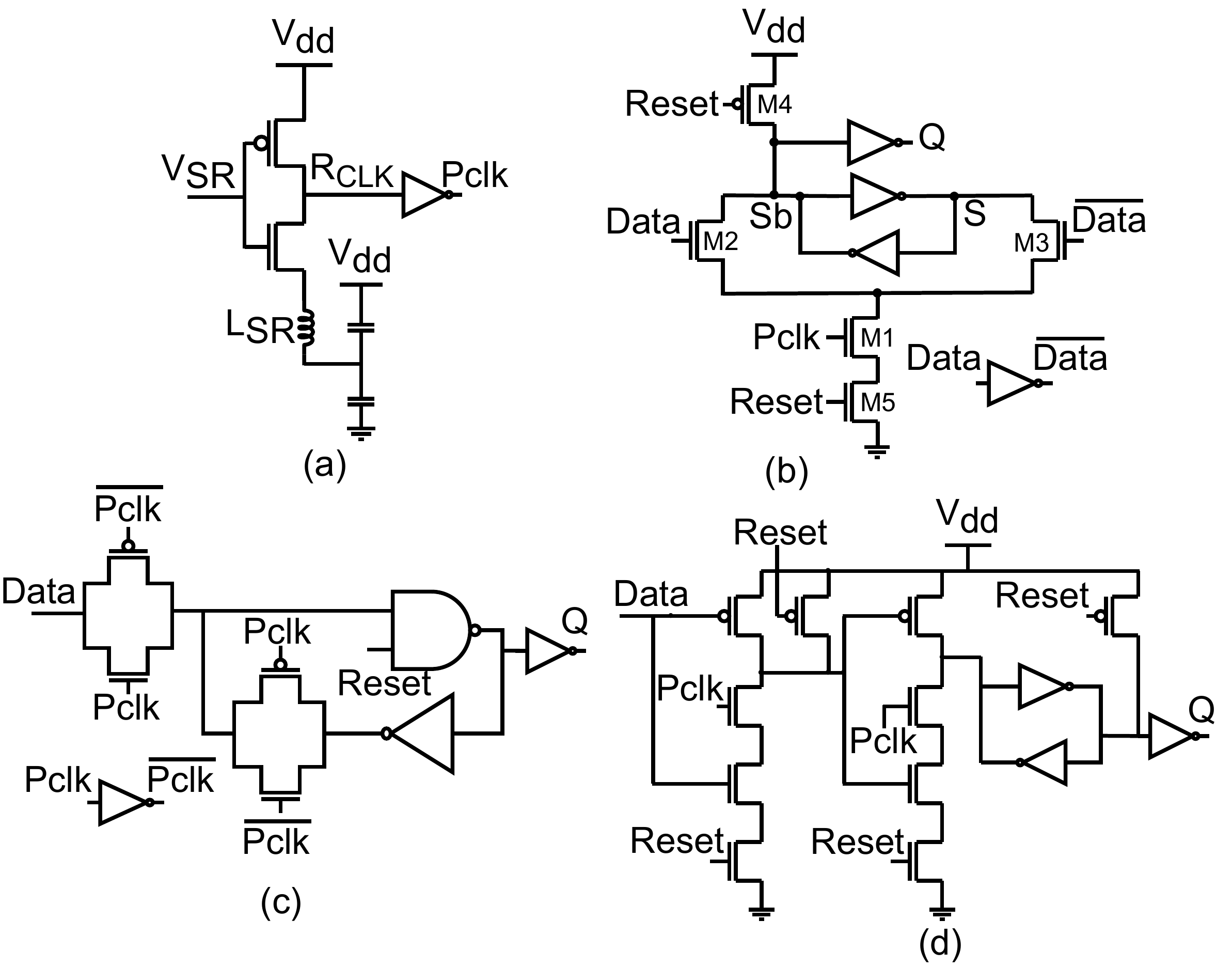}}
	\caption {The proposed series resonant pulsed FFs use (a) A PSR to generate pulse signal to drive the register stage, (b) 13T register, (c) pulsed register, (d) TSPC register to implement three pulsed FFs.} 
	\label{fig:all_reg}
	\vspace{-0.25cm}
\end{figure} 
\subsection{Proposed Resonant Pulsed Flip-Flops}
To utilize the generated $Pclk$ pulse (from the PSR, as shown in Fig.~\ref{fig:all_reg}(a)), we propose a 13T pulsed FF (13TPFF), as shown in Fig.~\ref{fig:all_reg}(b). It takes the input data and inverts it to provide it to the transistors M2 and M3, respectively. 
The M2 and M3 transistors drain are connected to the storage cells where the data is stored as logic ``1'' or a logic ``0''. If a ``1''  is stored in the register, the value at $S=1$ and $S_B=0$. If a ``0'' is stored, the voltages will be reversed.

When $Pclk$ is ``0,'' the transistor M1 is turned off, wherein the FF is in hold/retain state, and the values of $S$ and $S_B$ are unaltered. Consider the case when $Data=1$ and $Pclk=1$, the transistors M2 and M1 turn on connecting the node $S_b$ to ground, which then discharges the node and makes it 0, making $Q=1$ writing a ``1'' into the register. When $Data=0$ and $Pclk=1$, the transistors M3 and M1 turn ``ON" and write a ``0" at node $Q$.
The FF has an active-low asynchronous reset. 
The M4 and M5 transistors are turned ``ON'' and ``OFF,'' respectively, when the $Reset$ signal goes to low resulting in a logic ``1'' at node $S_b$ that writes a ``0'' the output $Q$.

Along with a 13TPFF, we also design pulsed energy recovery FFs. The pulsed resonant FF (PRFF) is based on the traditional latch, and the resonant TSPCFF is based on the TSPC register~\cite{Rabaey:2010}. Except, they use pulsed series resonance to take advantage of the input pulse signal $V_{SR}$ to recycle energy. The energy recovery FFs are positive edge-triggered with asynchronous active low $Reset$ signal. The use of conventional registers makes an easy integration of resonance clock trees into existing clock tree architectures.

\subsection{Skew Reduction Methodology}

Skew is defined as the spatial variation in the arrival time of clock transition at two different locations. There are several reasons for this skew, and one such cause is different loads on clock drivers~\cite{Rabaey:2010}.

In Fig. \ref{fig:testbench_clk_tree}, the eight different branches of the clock tree are having eight different capacitances $C_{SR1}$, $C_{SR2}$, and up to $C_{SR8}$ due to on-chip variation (OCV). This capacitance mismatch between different branches of a clock tree will result in different clock arrival times. 
Each branch of the clock tree represents a separate $LC$ resonant tank. Using the resonant frequency $f_{RES}=1/2\pi\sqrt{LC}$, we match the inductors $L_{SR1}$ to $L_{SR8}$ with the load capacitances $C_{SR1}$ to $C_{SR8}$, respectively, to have equal frequencies. This inductor matching would result in equal frequency signals in all the clock branches, thus, reducing the skew. The resonant frequency independent of the input clock frequency will not be affected by wide frequency band operation. The primary reason is the $delay = T_d$ of the pulse generator circuit is independent of clock pulse width, and it works on the clock edges. For all the clock frequencies less than the resonant frequency $f_{RES}$, we can have the same inductor value that results in reduced skew. Our results in Section~\ref{subsec:clock_tree} also supported this claim.

\section{ Results and Discussion }

\subsection{Experimental Setup}

\begin{figure*}
\captionsetup{aboveskip=-0.00cm,belowskip=-0.250cm}
    \centering
    \vspace{-0.5cm}
    \subfigure[]{\includegraphics[width=0.22\textwidth]{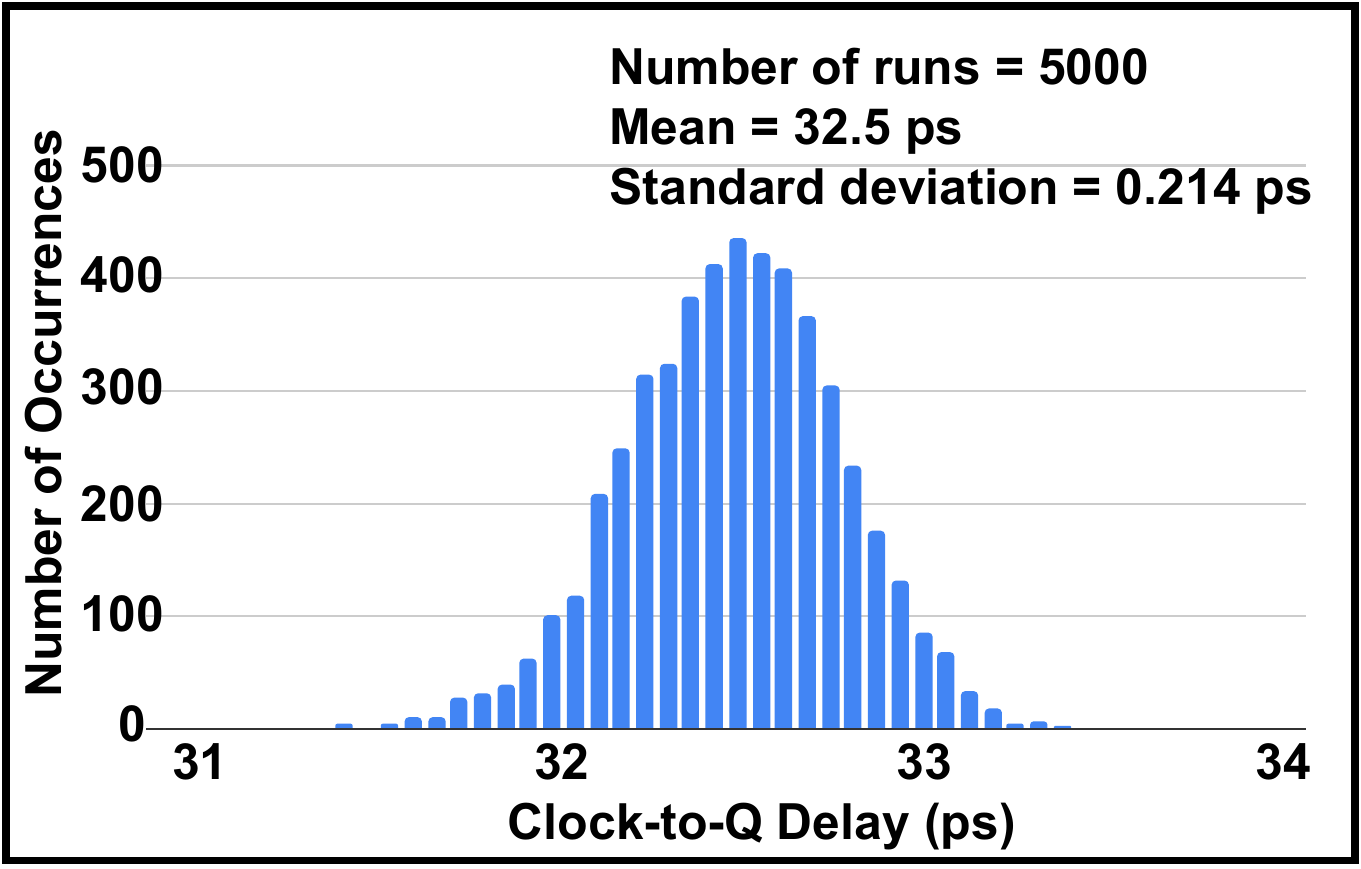}} 
     \subfigure[]{\includegraphics[width=0.22\textwidth]{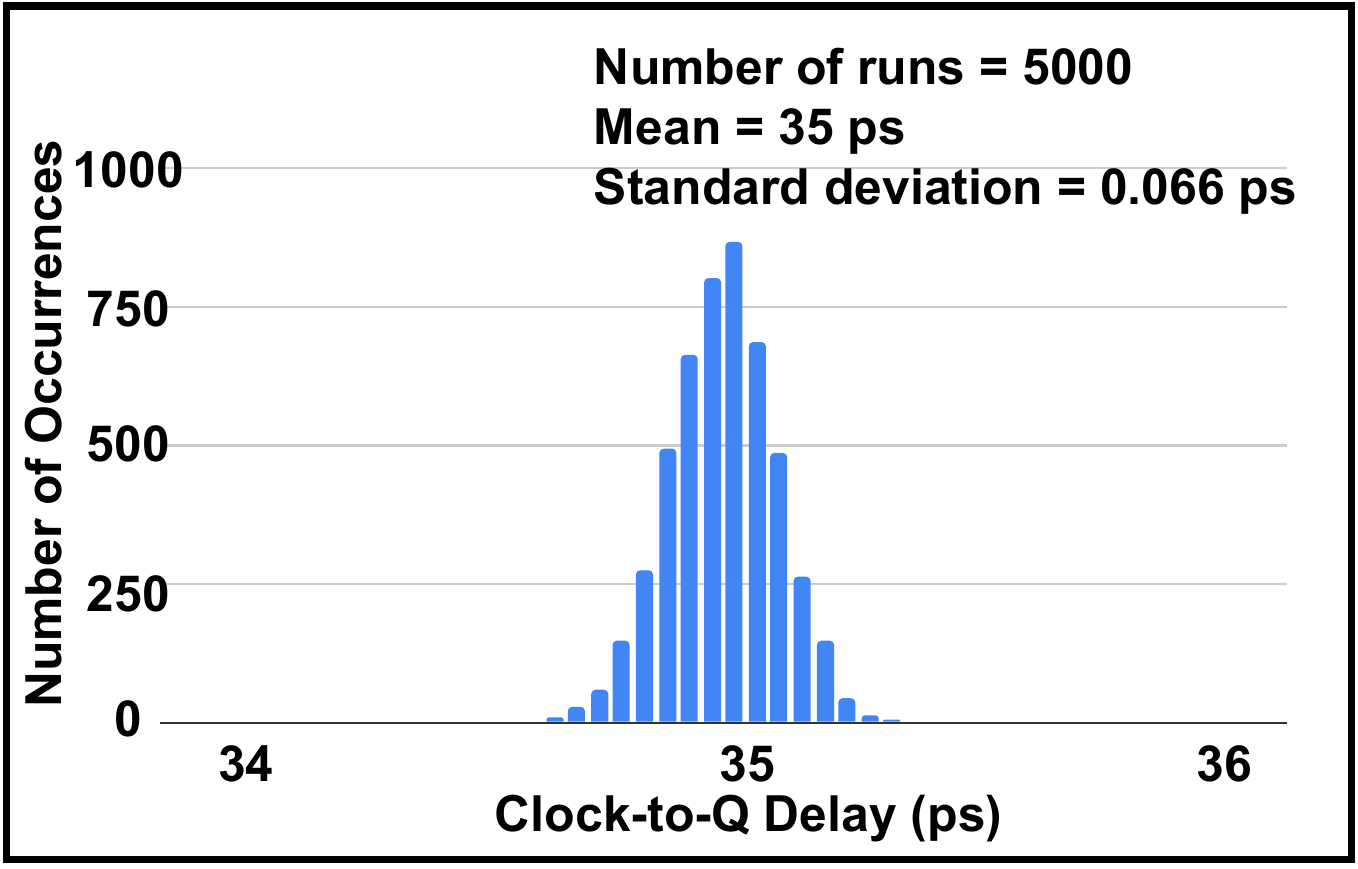}} 
    \subfigure[]{\includegraphics[width=0.22\textwidth]{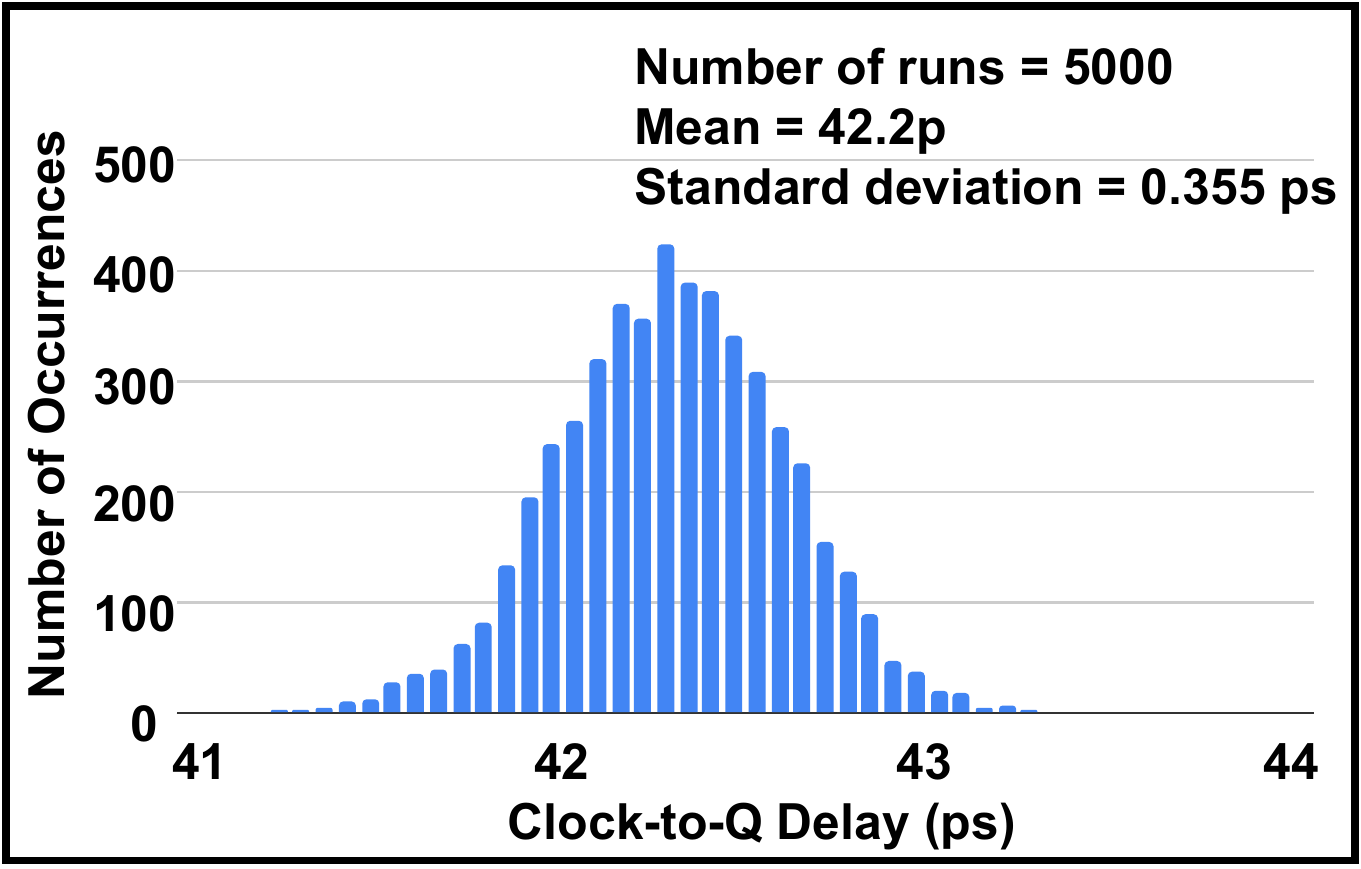}} 
    \subfigure[]{\includegraphics[width=0.22\textwidth]{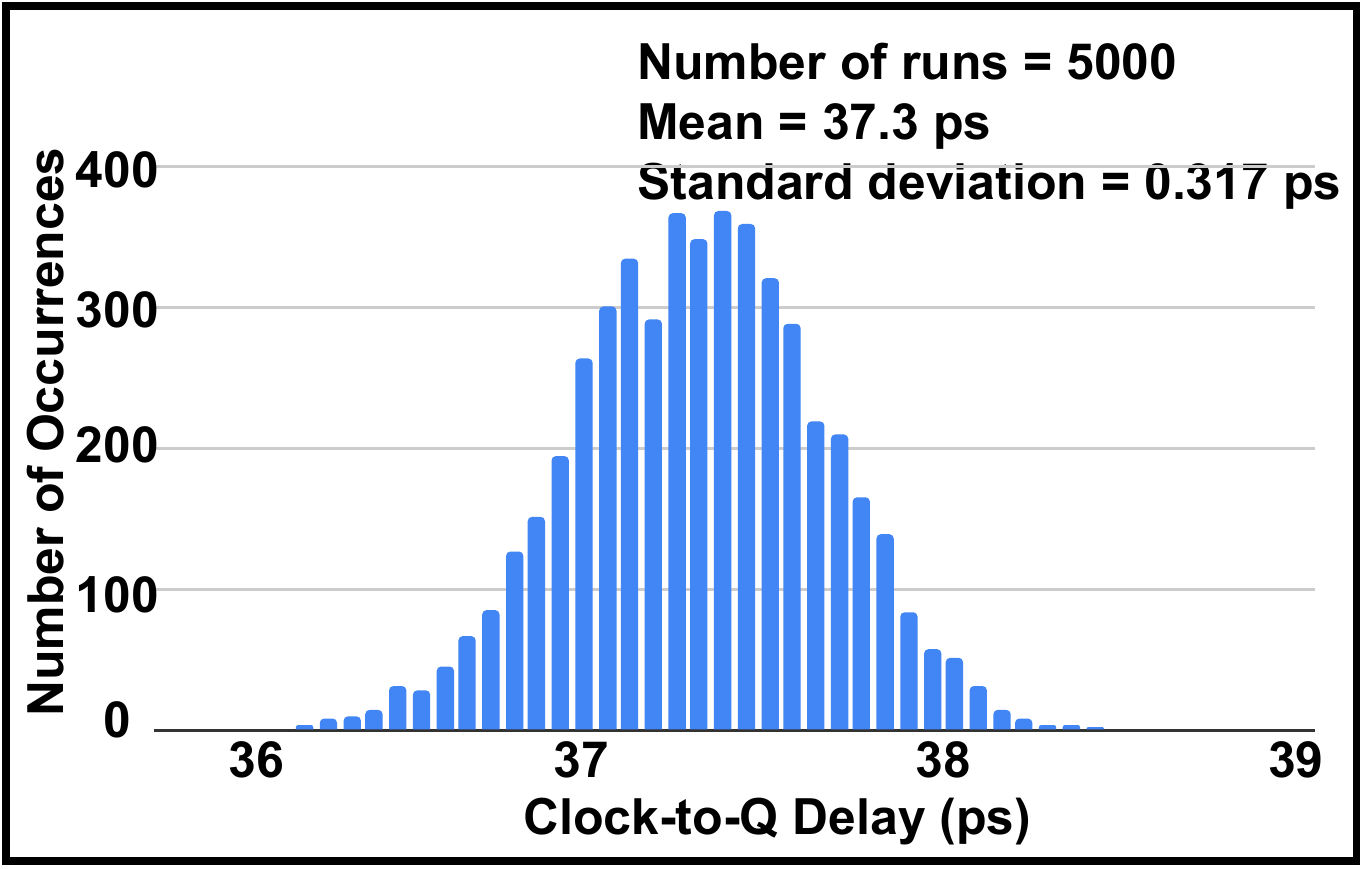}}
    \caption{Illustration of Monte-Carlo simulation results for various FFs by considering 5000 samples with $\pm 10\%$ length variation. (a) PSFF, (b) PRFF, (c) TSPCFF, and (d) 13TPFF.}
    \label{fig:monte_sims}
    \vspace{-0.50cm}
\end{figure*}

The proposed resonant clock tree architecture, as shown in Fig.~\ref{fig:testbench_clk_tree} is implemented using a standard 14nm FinFET technology. 
Conventional clock tree architecture is used as a reference model to compare with the proposed architecture.
Each tree has eight clock drivers, 4K clock gaters, 8K clock buffers, and 32K FFs. The traditional clock tree makes use of transmission gate PSFFs, whereas the resonant clock tree uses pulsed FFs, as shown in Fig.~\ref{fig:testbench_clk_tree}(a) and Fig.~\ref{fig:testbench_clk_tree}(b), respectively. 
All the FFs layouts are compatible with a standard cell height of 24 horizontal M2 tracks. All the simulations are performed for frequencies ranging from 1~${GHz}$ to 5~${GHz}$.
\begin{figure}[t!]
	\centerline{\includegraphics[width = 0.48\textwidth]{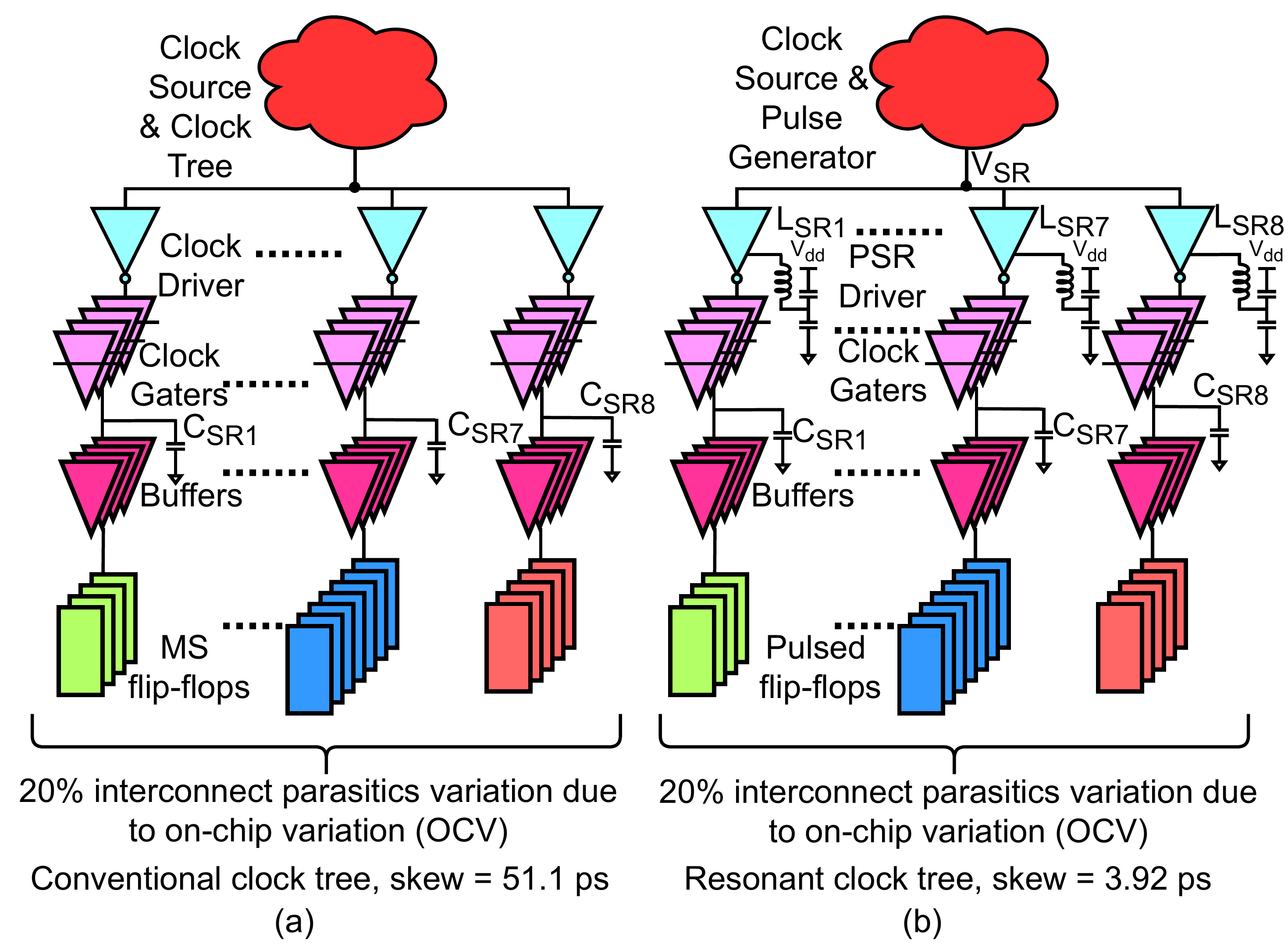}}
	\caption {Clock tree architectures used for functional simulations, (a) conventional clock tree architecture with eight branches and different loads totaling 32k FFs, (b) resonant clock tree architecture replicating the same number of branches and loads as the conventional one.} 
	\label{fig:testbench_clk_tree}
	\vspace{-0.25cm}
\end{figure} 

\begin{figure}[t!]
	\centerline{\includegraphics[width = 0.35\textwidth]{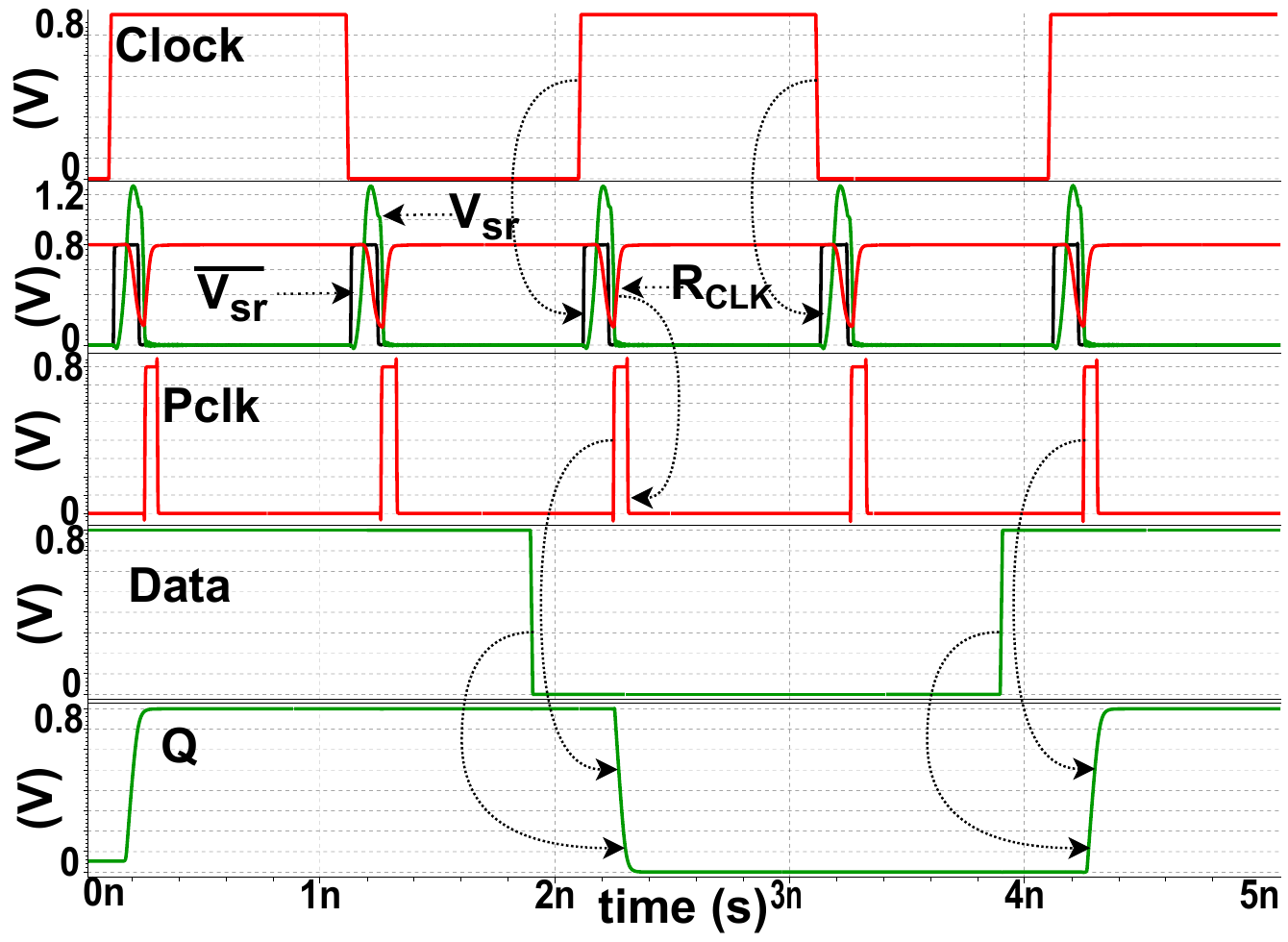}}
	\caption {Simulation waveforms show a CLK input of 0.5~${GHz}$ is provided to generate a 1~${GHz}$ $Pclk$ clock to assert as a clock input to the pulsed FFs.} 
	\label{fig:functionality_sim}
	\vspace{-0.25cm}
\end{figure} 

\begin{figure}[t!]
	\centerline{\includegraphics[width = 0.32\textwidth]{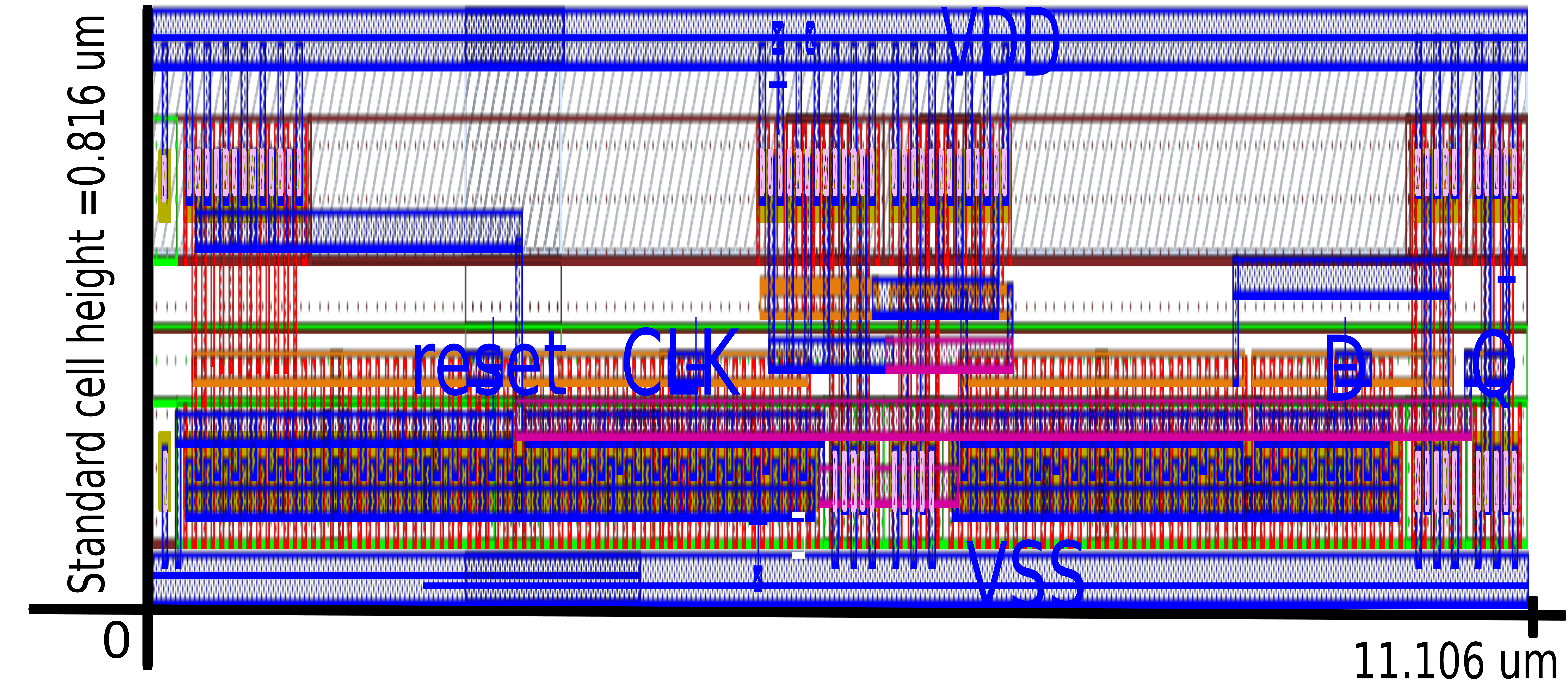}}
	\caption {13TPFF layout is implemented using 14 nm FinFET technology following standard cell height, which improves setup-time constraints compared to PSFF.} 
	\label{fig:layout_13t}
	\vspace{-0.5cm}
\end{figure} 

\begin{table}
\caption{The proposed 13TPFF exhibits better set-up time than the PSFF and better hold-time than TSPCFF and PRFF while consuming more dynamic power and area; however, it consume lower static power than PSFF and enables power saving in overall clock architecture.}
\centering
\label{tab:reg_comps}
\resizebox{\linewidth}{!}{%
\begin{tabular}{|c|c|c|c|c|c|c|c|c|c|c|c|} 
\hline
\multirow{2}{*}{\begin{tabular}[c]{@{}c@{}}\textbf{Types of}\\\textbf{FF}\end{tabular}} & \multirow{2}{*}{\begin{tabular}[c]{@{}c@{}}\textbf{Normalised}\\\textbf{area}\end{tabular}} & \multicolumn{3}{c|}{\textbf{Delay ($ps$)}} & \multicolumn{2}{c|}{\textbf{Static power ($pW$)}} & \multicolumn{5}{c|}{\textbf{Dynamic power ($\mu W$)}} \\ 
\cline{3-12}
 &  & \textbf{C-Q} & \textbf{ts} & \textbf{th} & \textbf{D=0} & \textbf{D=1} & \textbf{1GHz} & \textbf{2GHz} & \textbf{3GHz} & \textbf{4GHz} & \textbf{5GHz} \\ 
\hline
\textbf{PSFF} & 1 & 32.5 & 14 & 2 & 1550 & 593 & 8.3 & 14.1 & 21 & 28 & 35.1 \\ 
\hline
\begin{tabular}[c]{@{}c@{}}\textbf{PRFF}\end{tabular} & 0.59 & 35.1 & -95 & 96 & 278 & 272 & 7.16 & 13.8 & 20.4 & 27.1 & 33.8 \\ 
\hline
\begin{tabular}[c]{@{}c@{}}\textbf{TSPCFF}\end{tabular} & 0.84 & 41.9 & -92 & 93 & 283 & 664 & 12.3 & 20.2 & 28 & 35.9 & 43.7 \\ 
\hline
\begin{tabular}[c]{@{}c@{}}\textbf{13TPFF}\end{tabular} & 1.75 & 37.3 & -25 & 60 & 501 & 538 & 16.2 & 31.1 & 46 & 61 & 76 \\
\hline
\end{tabular}
}
\vspace{-0.5cm}
\end{table}

\begin{table}[ht]
\caption{Our proposed PRFF outperforms all the FFs and consumes 43\% less power than the conventional PSFF with ${11\times}$ improvement in skew, while TSPCFF and 13TPFF consume 26\% and 20.2\% less power than the PSFF, respectively.}
\label{tab:tree_power}
\centering
\resizebox{\linewidth}{!}{%
\begin{tabular}{|c|c|c|c|c|c|c|} 
\hline
\multirow{2}{*}{\begin{tabular}[c]{@{}c@{}}\textbf{Types of}\\\textbf{flip-flop}\end{tabular}} & \multirow{2}{*}{\textbf{Skew ($ps$) }} & \multicolumn{5}{c|}{\textbf{Total  tree power ($mW$)}} \\ 
\cline{3-7}
 &  & \textbf{1GHz} & \textbf{2GHz} & \textbf{3GHz} & \textbf{4GHz} & \textbf{5GHz} \\ 
\hline
\textbf{PSFF}~ & 51.1~ & 30.8~ & 60.6 & ~89.2 & ~116~ & 138 \\ 
\hline
\textcolor[rgb]{0.125,0.129,0.141}{\textbf{PRFF}} & 4.61 & 17.4 & 34.1 & 50.3 & 65.7 & 78.7 \\ 
\hline
\textcolor[rgb]{0.125,0.129,0.141}{\textbf{TSPCFF}} & 2.05 & 22.2 & 43.6 & 64.6 & 84.4 & 102 \\ 
\hline
\textcolor[rgb]{0.125,0.129,0.141}{\textbf{13TPFF}} & 3.92 & 23.8 & 46.8 & 69.4 & 90.7 & 110 \\
\hline
\end{tabular}
}
\vspace{-0.25cm}
\end{table}

\subsection{Implementation Results}

\subsubsection{Power and Performance Comparison of Registers}
For measuring the performance and functionality of the proposed 13TPFF under process variations, we consider 5000 samples of CLK-to-Q ($t_{c-q}$) delay using Monte-Carlo simulation. $\pm 10$\% variation in the length of all devices is considered while performing the simulations. The $t_{c-q}$ delay distributions of PSFF, PRFF, TSPCFF, and 13TPFFs are shown in Fig.~\ref{fig:monte_sims}(a),  Fig.~\ref{fig:monte_sims}(b), Fig.~\ref{fig:monte_sims}(c), and Fig.~\ref{fig:monte_sims}(d), respectively. Among all the resonant FFs, the PRFF has lowest mean $t_{c-q}$ of 35 ps with standard deviation of 0.066~$ps$.

The normalized layout area, $t_{c-q}$, setup times~$(t_s)$, hold times~$(t_h)$, and power for the FFs are listed in Table~\ref{tab:reg_comps}. Among all the competing FFs, the 13TPFF consumed the highest layout area of 9.62~${um^2}$, which is 1.75$\times$ the area of PSFF whose area is 5.151~${um^2}$, and 2.9$\times$ the area of a PRFF whose area is 3.091~${um^2}$. The proposed 13TPFF has a $t_s$ of -25~$ps$ and a $t_h$ of 60~$ps$ with a clock-to-q delay of 37.3~$ps$. Empirically, pulsed register-based FFs exhibit negative  $t_s$, which tremendously impacts resolving $t_s$ related timing issues. The PSFF has a $(t_s)$ of 14~$ps$, $(t_h)$ of 2~$ps$ and $t_{c-q}$ of 32.5~$ps$. The resonant PRFF has a better $(t_s)$ of -95~$ps$ but has a high $(t_h)$ of 96~$ps$ which is similar to the TSPCFF with -92~$ps$ $(t_s)$ and 93~$ps$ $(t_s)$. However, the power consumed by the proposed 13TPFF is 2$\times$ higher than the PSFF. Among all the competing FFs, the PRFF consumes the lowest dynamic static and dynamic powers.

\subsubsection{Clock Tree}
\label{subsec:clock_tree}

The functional simulation for the resonant clock is shown in Fig.~\ref{fig:functionality_sim}. We provide 0.5~${GHz}$ clock as input clock source shown in Fig.~\ref{fig:proposed_tree}. The output of the pulse generator is a boosted signal $V_{SR}$ of 1~${GHz}$ frequency. This $V_{SR}$ signal is then provided to a PSR driver whose output is $R_{CLK}$. 
This clock signals $IN_{1}$ along with the Data signal generates the output $Q$, as shown Fig.~\ref{fig:all_reg}(b). We compare the power and skew of the proposed clock architecture, as shown in Table~\ref{tab:tree_power} for frequencies ranging from 1~${GHz}$ to 5~${GHz}$. The power consumed by the proposed architecture while using 13TPFFs at 1~${GHz}$ frequency is 23.8~$mW$, compared to a conventional clock tree architecture with PSFF that consumes 30.8~$mW$. 
The skew generated by the conventional clock tree is 51.1~$ps$. As a result of inductor tuning, the skew generated by the proposed resonant clock architecture is 3.92~$ps$. The proposed architecture saves 22.7\% power using the 13TPFFs compared to conventional tree. The resonant clock tree using TSPCFFs has a skew of 2.1~$ps$ and saves 27.9\% power, whereas the clock with PRFFs has a skew of 4.6~$ps$ and saves 43\% power compared to conventional clock. 

\section{ Conclusion }

This paper proposed a resonant clock architecture to balance the skew and recycle the power consumed. The proposed architecture with 13TPFF saves 22.7\% power with 92\% lower clock skew than the conventional clock tree architecture. Furthermore, it saves 43\% power with a 91\% skew reduction while using the PRFF compared to conventional PSFF-based CMOS clock tree architecture in 14 nm FinFET technology.

\bibliographystyle{IEEEtran}
\bibliography{main}

\end{document}